\documentclass[reprint,aps, prl,amsmath,amssymb,amsfonts,superscriptaddress]{revtex4-2}

\usepackage{graphicx}
\usepackage{multirow}
\usepackage{color}
\usepackage{amsthm}
\usepackage{here}
\usepackage{braket}
\usepackage{bbold}

\newcommand{\HS}[1]{\textcolor{black}{{#1}}}

\begin{document}
\title{Partition of Two Interacting Electrons by a Potential Barrier}
\author{Sungguen Ryu}
\affiliation{Department of Physics, Korea Advanced Institute of Science and Technology, Daejeon 34141, Korea}
\affiliation{Institute for Cross-Disciplinary Physics and Complex Systems IFISC (UIB-CSIC), E-07122 Palma de Mallorca, Spain}
\author{H.-S. Sim} \email[]{hssim@kaist.ac.kr}
\affiliation{Department of Physics, Korea Advanced Institute of Science and Technology, Daejeon 34141, Korea}

\date{\today}

\begin{abstract}
Scattering or tunneling of an electron at a potential barrier is a fundamental quantum effect. Electron-electron interactions often affect the scattering, and understanding of the interaction effect is crucial in detection of various phenomena of electron transport and their application to electron quantum optics. We theoretically study the partition and collision of two interacting hot electrons at a potential barrier. We predict their kinetic energy change by their Coulomb interaction during the scattering delay time inside the barrier.
The energy change results in characteristic deviation of the partition probabilities from the noninteracting case. 
The derivation includes nonmonotonic dependence of the probabilities on the barrier height, which qualitatively agrees with recent experiments, and reduction of the fermionic antibunching.
\end{abstract}
\maketitle

Interplay of potential scattering of an electron and electron-electron interactions causes nontrivial effects.
Generally, the former is used for detecting the latter.
For example, the interaction strength of Luttinger liquids~\cite{Kane:ZBA} and spatial ordering of Wigner crystals~\cite{Glazman:ZBA,Ilani:wigner} are read out from anomalous electron tunneling through a potential barrier.
And, the latter reduces quantum coherence of the former.
It happens in electron interferometers in the quantum Hall regime~\cite{ji2003electronic,neder:unexpected,Litvin:MZI,roulleau:finite}, where phase accumulation between scattering events is smeared out by 
intra-~\cite{youn:nonequilibrium,Neder:Ginossar,Kovrizhin:MZI} or inter-edge-channel interactions~\cite{Neder:controlled,Levkivskyi:Sukhorukov}.

The interplay has been investigated in electron quantum optics.  Electron scattering at a potential barrier provides a tool not only for studying partition~\cite{henny1999fermionic,oliver1999hanbury}, antibunching~\cite{liu1998}, identical particle statistics and anyon braiding~\cite{BLee;FQHE,bartolomei:fractional,JYLee;IQHE,Mora,JYLee;non-Abelian},
but also for operating flying qubits~\cite{bauerle:coherent}. 
It combines with on demand generation of wave packets by AC driving~\cite{pekola:single,feve:demand,moskalets2008quantized, keeling:minimal,keeling2008coherent,dubois:minimal,giblin2012towards,Hohls,kaestner:nonadiabatic,ryu:ultrafast,yamahata:picosecond,Freise:counting,Hermelin2011,McNeil2011,Takada2019,Haug2021}.
An electron packet, generated on a quantum Hall edge at the Fermi level, is partitioned at a barrier. Using partition noise~\cite{blanter:shot}, one studies antibunching between the electron and excitations of the Fermi sea~\cite{bocquillon:electron}.
When two packets collide~\cite{ol:shot,jullien:quantum} at a barrier as in Hong-Ou-Mandel effects, deviation from fermionic antibunching was observed~\cite{bocquillon:coherence,freulon:hong} and attributed to charge fractionalization~\cite{grenier2013fractionalization,bocquillon2013separation,wahl2014interactions} of Luttinger liquids.

\begin{figure}[b]
 \includegraphics[width=.85\columnwidth]{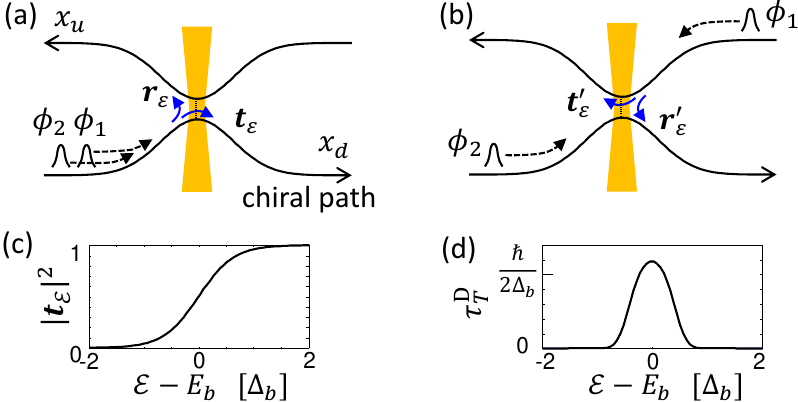}
 \caption{Potential barrier (shade) on chiral paths $u$ and $d$. (a) Partition of two copropagating hot electrons $\phi_{i=1,2}$ (peaks) at the barrier. (b) Collision of two counterpropagating electrons at the barrier. (c) Transmission probability $|\mathbf{t}_{\mathcal{E}}|^2$  and (d) delay time $\tau^\textrm{D}_T$ of a plane wave of energy $\mathcal{E}$ at the barrier. 
}
\label{fig:setup}
\end{figure}
 
All the above examples involve interaction effects outside a barrier. A recent experiment~\cite{ubbelohde:partitioning} 
implies interactions {\em inside} a barrier. There, two single-electron wave packets are generated far above ($\gtrsim$ 100 meV) the Fermi level by a quantum-dot pump.
These hot electrons copropagate in a depleted region, spatially isolated from other electrons. The observed partition probabilities of the electrons at the barrier [Fig.~\ref{fig:setup}(a)] cannot be decomposed into products of single-electron partition probabilities. 
The probabilities show nonmonotonic dependence on the barrier height.
The results are not described by noninteracting theories~\cite{Landauer,LandauerButtiker,hassler:wavepacket} nor by the charge fractionalization. 
They remain unexplained, suggesting that the characteristics of the barrier needs to be counted. 

In this Letter, we develop a scattering theory of two 
interacting hot electrons at a potential barrier, and notice a central role of the scattering delay times (sometimes called phase times~\cite{hauge:tunneling})
\begin{equation}
\tau^{\text{D}}_\textrm{T} \equiv  \hbar\, \text{Im}\, \frac{d\, \text{ln}\, \mathbf{t}_{\mathcal{E}}}{d \mathcal{E}}, \quad \quad \quad \tau^{\text{D}}_\textrm{R} \equiv  \hbar\, \text{Im}\, \frac{d\, \text{ln}\, \mathbf{r}_{\mathcal{E}}}{d \mathcal{E}} \label{scattering_delay_time}
\end{equation}
of single-electron transmission and 
reflection at the barrier.
$s_{\alpha \beta}(\mathcal{E}) \in \{ \mathbf{r}_{\mathcal{E}},\mathbf{r}'_{\mathcal{E}}, \mathbf{t}_{\mathcal{E}}, \mathbf{t}'_{\mathcal{E}}\}$ is the scattering amplitude of a plane wave of energy $\mathcal{E}$ from an input path $\beta$  to an output $\alpha$ at the barrier [Fig.~\ref{fig:setup}].
We predict the kinetic energy change of the electrons by their Coulomb interaction during the delay times,
and compute its effect on their partition at the barrier, considering initially copropagating or counterpropagating electrons. 
In the copropagating case,
our theory explains the recent experiments~\cite{ubbelohde:partitioning}.
Energy dependence of the delay times causes nonmonotonic dependence of the partition on the barrier height. 
The scattering probabilities of the two electrons are correlated when the transmission and reflection delay times differ. 
In the counterpropagating case, we distinguish direct and exchange interaction effects on the partition, especially on the reduction of their antibunching.
 



{\it Setup.---} We consider two hot electrons generated by quantum-dot pumps in \HS{a strong magnetic field}~\cite{kataoka:time,Freise:counting}.
They approach a potential barrier, propagating along a one-dimensional chiral upper path $\gamma = u$ or a lower path $\gamma = d$  in depleted regions. 
In Fig.~\ref{fig:setup}(a), they initially copropagate, occupying orthogonal single-electron wave packets $\phi_{m=1,2}$ which usually separate in energy or time in experiments~\cite{ubbelohde:partitioning}.
In Fig.~\ref{fig:setup}(b), they initially counterpropagate, occupying packets $\phi_{m=1,2}$ of the same Gaussian form~\cite{ryu:ultrafast}, and arrive at the barrier simultaneously. 
Each initially has kinetic energy $E^{(0)}_m$ and energy uncertainty $\sigma_E$.
We assume that their propagation velocity $v$ is energy independent, as the dependence is not strong enough to generate the nonmonotonicity~\cite{ubbelohde:partitioning}. 

\HS{In the strong magnetic field}, the barrier is described by a saddle point constriction~\cite{buttiker:quantized}, and mapped onto a one-dimensional problem~\cite{fertig:transmission}.
For a plane wave of energy $\mathcal{E}$, the barrier transmission probabilities, 
$|\mathbf{t}_{\mathcal{E}}|^2 =|\mathbf{t}'_{\mathcal{E}}|^2= 1/[1+\exp(-\pi(\mathcal{E}-E_b)/\Delta_b)]$,
change from 0 to 1 over the energy $\Delta_b$
around the barrier height $E_b$ where $|\mathbf{t}_{\mathcal{E}=E_b}|^2 = 0.5$
 [Fig.~\ref{fig:setup}(c)].
We consider the $\sigma_E < \Delta_b$ regime to predict universal results; here, the wave packet form does not change during its barrier scattering, hence, the results are insensitive to the form.
In Ref.~\cite{ubbelohde:partitioning},  $\Delta _b \sim 5 \sigma_E$.
 
The electrons interact through a Coulomb potential~\cite{bellentani:coulomb}, $W(x_{\text{rel}}) = W_0 e^{- x_{\text{rel}} /a_{\text{scr}}} / \sqrt{1 + (x_{\text{rel}}/a_{\text{cut}})^2}$. 
Their separation $x_{\text{rel}}$ is simplified as $x_{\text{rel}} = x_{1} -x_{2}$
when their coordinates $x_{m}$ are on the same path,
and $x_{\text{rel}} = |x_{1}| + |x_{2}|$  
for them on different paths ($x_m=0$ at the barrier). 
$a_{\text{scr}}$ is the screening length.
The cut-off $a_{\text{cut}}$ describes packet broadening to the transverse directions by the magnetic length or the quantum well width confining two-dimensional electrons.
 
{\it Interaction during delay times.---} 
We compute the partition probabilities $P_{n}$ that $n$ ($=0,1,2$) of the two electrons move to the lower path after barrier scattering. They have contributions $P_{n}  = P_n^{(\text{dir})}  +P_n^{(\text{ex})}$ from direct and exchange processes, $P_n^{(\text{dir})} =  \braket{\hat{P}_n}$, $P_n^{(\text{ex})} = \mp \braket{\hat{P}_n \mathcal{P}_{\text{ex}}}$.
$\hat{P}_n$ is the projection operator onto the event of $P_n$.  $\mathcal{P}_{\text{ex}}$ is the operator exchanging the two electrons. The sign $-$ ($+$)  is for the electrons in the spin triplet (singlet).
We obtain~\cite{supple} the correction $\delta P_n^{\text{(dir/ex)}}$ to the noninteracting probabilities $P_n^{(0)}$  up to the lowest order of the interaction $W$ and $\sigma_E/\Delta_b$, 
\begin{equation}
\label{eq:P-1st-dir} 
\begin{aligned}
  &\delta P_n^{\text{(dir)}} = 
  \int_0^\infty dt (- \frac{i}{\hbar} ) \braket{
    [\hat{P}_n  , W] }
    _{\phi_{1}(t)\otimes \phi_{2}(t)}     
  \\
  &\delta P_n^{\text{(ex)}}
    = \mp 
    \int_0^\infty \hspace{-0.1cm} dt  (-\frac{i}{\hbar} ) \braket{
    [ \hat{P}_n \mathcal{P}_{\text{ex}} , W] }
    _{\phi_{1}(t)\otimes \phi_{2}(t)}   
\end{aligned}
\end{equation}
by perturbatively expanding the time evolution operator with respect to $W$.
$[\cdots,\cdots]$ is the commutator.

$\phi_{m=1,2}(t)$ are the packets at time $t$ in the noninteracting case.
Their product state is used in computing the expectation values $\langle \cdots \rangle$ in Eq.~\eqref{eq:P-1st-dir}, assuming that the packets are separable at the initial time $t=0$.
Each is decomposed
into $\phi^\text{(in)}_m$ in the input path $\beta$, $\phi^\text{(out, $\alpha$)}_m$ in an output path $\alpha$, and $\phi^\text{(bar)}_m$ in the barrier,
\begin{equation} \label{eq:phi-expand}
\ket{\phi_{m}(t)} = \ket{{\phi}^\text{(in)}_{m}(t)} +  \sum_{\alpha=u,d} \ket{{\phi}^{(\text{out},\alpha)}_{m}(t)} + \ket{\phi^{\text{(bar)}}_{m}(t)}.
\end{equation}
The expression $\ket{{\phi}^{(\text{out},\alpha)}_{m}(t)}$ includes the scattering amplitude $s_{\alpha \beta}$.
For $\sigma_E \ll \Delta_b$,
we derive~\cite{supple} 
the probability of electron $m \, (=1,2)$ being in the barrier, 
\begin{equation}
  \label{eq:prob-barr}
  \braket{\phi_{m}^{\text{(bar)}} (t)|\phi_{m}^{\text{(bar)}}(t)}
  =   \bar{\tau}_m  A_m(t) + \mathcal{O}(\sigma_E^2 / \Delta_b^2),
\end{equation}
in terms of the barrier dwell time~\cite{hauge:tunneling} (mean delay time) $\bar{\tau}_m \equiv |\mathbf{t}_{E_m^{(0)}}|^2 \tau^{\text{D}}_{m \text{T}} +|\mathbf{r}_{E_m^{(0)}}|^2 \tau^{\text{D}}_{m \text{R}}$ 
and the arrival time distribution~\cite{emary:phonon} $A_m(t)$ (the probability per time of arrival at the barrier at $t$) of electron $m$. 
  
  

Inside the barrier, electron $m$  has the kinetic energy $E_m = E_m^{(0)} + \delta E_{m}^\textrm{(dir)} + \delta E_{m}^\textrm{(ex)}$.
The change $\delta E_{m}^\textrm{(dir/ex)}$ from the initial value $E_m^{(0)}$
by  direct/exchange interactions with the other electron $m'$ occurs in their input paths or the barrier, hence, depending on the trajectory of $m'$.
Using Eqs.~\eqref{eq:P-1st-dir}-\eqref{eq:prob-barr}, the energy change occurring in the barrier during the dwell time $\bar{\tau}_m$ of $m$ is found~\cite{supple} as $\bar{\tau}_m  \Gamma^\textrm{(dir/ex)}+\mathcal{O}(\sigma_E^2 / \Delta_b^2) + \mathcal{O}(W^2)$,
\begin{equation}
\label{kinetic_direct}
\begin{aligned}
&  \bar{\tau}_m \Gamma^\textrm{(dir)}  =   -   \bar{\tau}_m \int dt  A_m (t)   \braket{v \frac{\partial W}{\partial  x_m} }_{\ket{0_m}\otimes \ket{\phi_{m' |\alpha}(t)}}, \\ 
& \bar{\tau}_m \Gamma^\textrm{(ex)}  =   -  \bar{\tau}_m \textrm{Re} \int dt  A_m(t)   \braket{  v
    \frac{\partial W}{\partial  x_m }   \mathcal{P}_{\text{ex}}}_{\ket{0_m}\otimes \ket{\phi_{m' |\alpha}(t)}}. 
\end{aligned}
\end{equation}
The Coulomb power $\Gamma^\textrm{(dir/ex)}$ comes from the force $-  \partial W / \partial x$ to electron $m$ while $m$ is inside the barrier (described by the state $\ket{0_m}$) and $m'$ moves along a trajectory from its input $\beta$ to output $\alpha$  without partitioning at the barrier that is described by $ \ket{\phi_{m'| \alpha}(t)}  \equiv \ket{\phi_{m'}^{(\text{in})}(t)} +\ket{\phi_{m'}^{(\text{barr})}(t)} + s^{-1}_{\alpha\beta} \ket{\phi_{m'}^{(\text{out},\alpha)}(t)}$ [cf. the corresponding state with partitioning in Eq.~\eqref{eq:phi-expand}].
%
%
%

The energy change $\delta E_{m}^\textrm{(dir/ex)}$ modifies
the partition probabilities. 
$P_2  = P_2^{(\text{dir})} + P^{(\text{ex})}_2$ is found as
\begin{equation}
\label{eq:Pdir}
\begin{aligned} 
   &P_2^{(\text{dir})}
   \simeq \prod_{m=1,2} \int d \mathcal{E} \, |\tilde{\phi}_m(\mathcal{E})|^2 |s_{d \beta_m}(\mathcal{E} + \delta E_{m}^{(\text{dir})})|^2,   \\ 
   &P^{(\text{ex})}_2
    \simeq \mp \Big|\int d \mathcal{E} \,
     \big(\tilde{\phi}_1(\mathcal{E})
      s_{d \beta_1}(\mathcal{E}+ \delta E^{(\text{ex})}_{1})\big)^*  \\
  & \qquad \qquad \qquad \qquad\times   \tilde{\phi}_2(\mathcal{E}) s_{d \beta_2}(\mathcal{E} + \delta E^{(\text{ex})}_{2})   \Big|^2
\end{aligned}
\end{equation}
with (i) the amplitude $\tilde{\phi}_m(\mathcal{E})$ of finding the initial packet of electron $m$ in the plane wave having energy $\mathcal{E}$ of its input path $\beta_m$ and (ii) the scattering amplitude $s_{d \beta_m}$ to the lower output $d$ at the energy shifted by $\delta E_{m}^{(\text{dir/ex})}$ due to the interaction with the other electron $m'$ moving to the lower output. 
Equation~\eqref{eq:Pdir} is valid up to the lowest order of $W$ and $\sigma_E/\Delta_b$,
and gives the noninteracting result at $\delta E_{m}^{(\text{dir/ex})} = 0$.  $P_0$ is found similarly for two electrons moving to the upper output
and $P_1 =1-P_0  -P_2$.
  




{\it Partitioning copropagating electrons.---}
We consider two copropagating electrons [Fig.~\ref{fig:setup}(a)],
the predecessor (labeled by $m=1$) and successor ($m=2$) initially separated by distance $\ell > \hbar v/(2\sigma_E)$. 
Their partition is determined by direct processes. 
When $\sigma_E \ll \Delta_b$,
the partition probabilities in Eq.~\eqref{eq:Pdir} are written as
\begin{equation}
\label{eq:P-HBT} 
\begin{aligned}
  P_2
  & \simeq 
  \big|\mathbf{t}_{\tilde{E}_1 + \delta E_{1|TT}^{\text{(dir)}}}\big|^2  \big|\mathbf{t}_{\tilde{E}_2 + \delta E_{2|TT}^{(\text{dir})}}\big|^2, \\
  P_0
  & \simeq  \big|\mathbf{r}_{\tilde{E}_1 +\delta E_{1|RR}^{\text{(dir)}}}\big|^2  \big|\mathbf{r}_{\tilde{E}_2 +\delta E_{2|RR}^{(\text{dir})}}\big|^2.
\end{aligned}
\end{equation}
In the noninteracting limit, they are $P_2  =|\mathbf{t}_{E_1^{(0)}}|^2 |\mathbf{t}_{E_2^{(0)}}|^2$ and $P_0 = |\mathbf{r}_{E_1^{(0)}}|^2 |\mathbf{r}_{E_2^{(0)}}|^2$. 
$\tilde{E}_m - E_m^{(0)}$ is the kinetic energy change of electron $m$ 
that happens while the electrons copropagate along the input path over distance $L_{m=1,2}$;  the predecessor gains energy, $\tilde{E}_1 - E_1^{(0)} = \Gamma_\ell L_1 /v$, and the successor losses energy, $\tilde{E}_2 - E_2^{(0)} = - \Gamma_\ell L_2 /v$. 
$\Gamma_\ell \equiv - v  \frac{\partial W}{\partial x_{\text{rel}}}|_{x_{\text{rel}}=\ell}$  ($>0$) 
is the Coulomb power at their separation $\ell$. The energy gain or loss is determined by 
the sign of the force $- \partial W / \partial x_m$ [cf. Eq.~\eqref{kinetic_direct}].
Electron $m$ has further energy change by $\delta E_{m|TT}^{(\text{dir})}$ (resp. $\delta E_{m|RR}^{(\text{dir})}$) during barrier scattering when they both are transmitted (resp. reflected).
We roughly estimate it from Eq.~\eqref{kinetic_direct},
\begin{equation}
  \label{eq:dE-HBT}
  \begin{aligned}
  \delta E_{1|TT}^{(\text{dir})} &\approx \Gamma_\ell \bar{\tau}_1, \quad  \,\,   \delta E_{2|TT}^{(\text{dir})}
    &\approx   -\Gamma_\ell \bar{\tau}_1
    -\Gamma_{\ell-v \tau^{\text{D}}_{1\text{T}}} \bar{\tau}_2,  \\
    \delta E_{1|RR}^{(\text{dir})} &\approx \Gamma_\ell \bar{\tau}_1, \quad  \,\,
    \delta E_{2|RR}^{(\text{dir})} 
    &\approx   -\Gamma_\ell \bar{\tau}_1
    -\Gamma_{\ell-v \tau^{\text{D}}_{1\text{R}}} \bar{\tau}_2.
  \end{aligned}
\end{equation}
During its dwell time $\bar{\tau}_1$  the predecessor
gains energy $\Gamma_\ell \bar{\tau}_1$, while the successor losses $\Gamma_\ell \bar{\tau}_1$.
After the predecessor scatters out of the barrier, the successor enters the barrier, as  $\ell > \hbar v/(2\sigma_E)$.  This moment, their separation is reduced to $\ell - v \tau^{\text{D}}_{1\text{T}}$ or  $\ell - v \tau^{\text{D}}_{1\text{R}}$ by the delay time of the barrier transmission or reflection of the predecessor.
Then the successor further losses energy by $\Gamma_{d-v \tau^{\text{D}}_{1\text{T}}} \bar{\tau}_2$ or $\Gamma_{d-v \tau^{\text{D}}_{1\text{R}}} \bar{\tau}_2$ during its dwell time $\bar{\tau}_2$.

\begin{figure}[t]
  \includegraphics[width=.9\columnwidth]{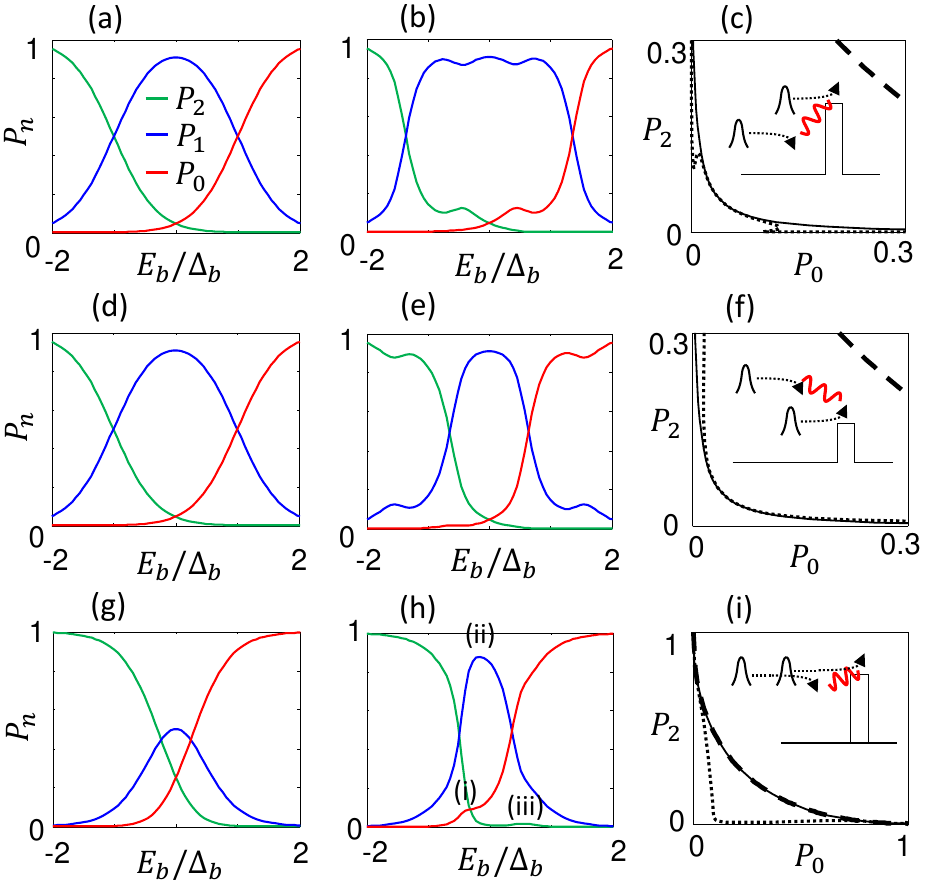}
  \caption{Partition probabilities $P_{n=0,1,2}$ of two copropagating electrons in Fig.~\ref{fig:setup}(a) by the symmetric saddle point constriction $V_{\text{sym}}$, as a function of the barrier height $E_b$ measured with respect to $(\tilde{E}_1 +\tilde{E}_2)/2$.
Left panels: The noninteracing case. 
Middle: The interacting case.
Right: $P_0$ versus $P_2$ in the noninteracting (solid curve) and interacting (dotted) cases.
The thick dashed curve follows $\sqrt{P_0}+\sqrt{P_2} = 1$.
Insets: Schematic kinetic energy change of the electrons during barrier scattering.
In (a)-(c), $\tilde{E}_1 = \tilde{E}_2 + 2 \Delta_b$.
In (d)-(f), $\tilde{E}_1  = \tilde{E}_2  - 2 \Delta_b$.
In (g)-(i), $\tilde{E}_1  =\tilde{E}_2$.
We choose 
$W_0= $ 144 meV~\cite{supple}, $a_{\text{scr}}= 500$ nm, 
$a_{\text{cut}}= 10$ nm, 
$\Delta_b = $ 5.4 meV~\cite{ubbelohde:partitioning}, $\sigma_E =$ 1 meV~\cite{ryu:ultrafast}, $v = 5 \times 10^{4}$ m/s~\cite{kataoka:time}, and $\ell = 3 \hbar v /(2\sigma_E)$.
 }
\label{fig:HBT_sQPC}
\end{figure}
\begin{figure}[t]
 \includegraphics[width=\columnwidth]{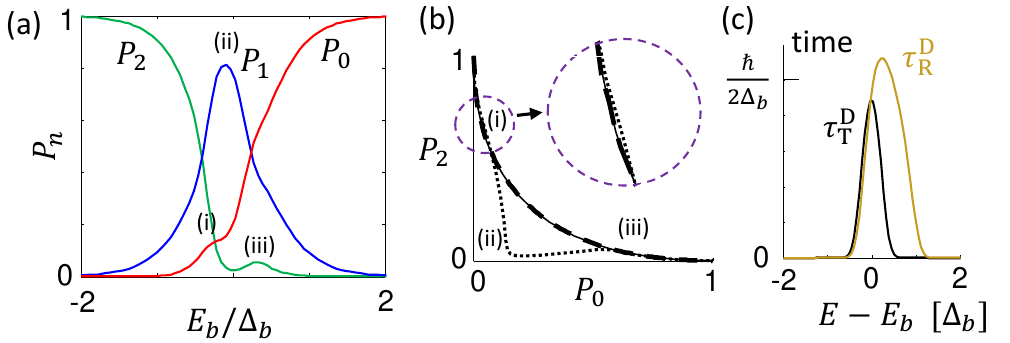}
 \caption{Partition probabilities $P_{n}$ of two copropagating electrons, having $\tilde{E}_1 =\tilde{E}_2$, by the asymmetric constriction $V_{\text{asym}}$.
   (a) $P_{n}$ as a function of $E_b$ in the interacting case [cf. the corresponding symmetric constriction in Figs.~\ref{fig:HBT_sQPC}(g)-(i)].
   (b) $P_0$ versus $P_2$ in the noninteracting  (solid curve) and interacting (dotted)  cases.
The interacting case violates $\sqrt{P_0}+\sqrt{P_2} \le 1$ (the dashed curve in the zoom-in plot). 
   (c) Delay times $\tau^{\text{D}}_{\text{T}}$ and $\tau^{\text{D}}_{\text{R}}$ for the transmission and reflection of a packet of energy $E$ at the asymmetric constriction.
The same parameters with Fig.~\ref{fig:HBT_sQPC} are chosen, except $\omega_{x \text{L}} / \omega_{x \text{R}} = 1/2$.
 }
\label{fig:HBT_asQPC}
\end{figure}

Using Eq.~\eqref{eq:Pdir}, we compute $P_n$ in Fig.~\ref{fig:HBT_sQPC} for a symmetric saddle point constriction  $V_{\text{sym}}=E_b -m^* \omega_0^2 (x^2 - y^2)/2$ on the two dimension $(x,y)$. 
The results qualitatively follow Eqs.~\eqref{eq:P-HBT} and \eqref{eq:dE-HBT}.
This constriction has  $\Delta_b =\hbar \omega_0^2/(2\omega_c)$~\cite{fertig:transmission}
and the symmetric delay times, $\tau^{\text{D}}_{m \text{T}} = \tau^{\text{D}}_{m \text{R}} = \bar{\tau}_m$, hence $\delta E_{2|TT}^{(\text{dir})}=\delta E_{2|RR}^{(\text{dir})}$. $\omega_c$ is the cyclotron frequency and $m^*$ is the electron effective mass.
The partition probabilities 
exhibit nonmonotonic dependence on $E_b$ in various energy configurations of $\tilde{E}_{m=1,2}$. 
This originates from the peak structure in the energy dependence of the delay times [Fig.~\ref{fig:setup}].
For instance, the energy exchange $\Gamma_\ell \bar{\tau}_1$  is maximal when the energy $\tilde{E}_1$ of the preceding electron aligns with the barrier height so that $\bar{\tau}_1$ is the largest.
\HS{The resulting nonmonotonic features of $P_n$ at $\tilde{E}_1, \tilde{E}_2 \sim E_b$,
the enhanced $P_1$ [see (ii) in Fig.~\ref{fig:HBT_sQPC}(h) and Fig.~\ref{fig:HBT_asQPC}(a)]
and the reduced $P_0$ and $P_2$ accompanied by peaks [(i) and (iii)],
agree with the corresponding features of Fig.~3d of the experimental report~\cite{ubbelohde:partitioning}.}

In an asymmetric saddle point constriction, the transmission and reflection delay times $\tau_\textrm{1T}^\textrm{D}$ and $\tau_\textrm{1R}^\textrm{D}$ differ. Then the partition can violate $\sqrt{P_0} +\sqrt{P_2} \le 1$, a condition~\cite{hassler:wavepacket} for uncorrelated scattering of noninteracting electrons.
To see this, we choose an asymmetric constriction $V_{\text{asym}} (x,y)= E_b -m^*( \omega_{x}^2 x^2 - \omega_{y}^2 y^2)/2$ where $\omega_x = \omega_{x \text{L}}$ and $\omega_{y} = \omega_{y \text{L}}$ for $x<0$,
$\omega_x = \omega_{x \text{R}}$ and $\omega_{y} = \omega_{y \text{R}}$ for $x>0$, and $\omega_{x \text{L}}/\omega_{y \text{L}}= \omega_{y \text{R}}/\omega_{x \text{R}}=1/2$, $\omega_{y \text{L}}=\omega_{x \text{R}}$;  the violation does not rely on this specific choice for simplicity of calculation.  It has $\Delta_b = \hbar \omega_{x \text{L}} \omega_{y \text{L}}/(2 \omega_c)$ and $\tau_\textrm{1T}^\textrm{D} < \tau_\textrm{1R}^\textrm{D}$~\cite{supple}.
Then the reflection of the predecessor, in comparison with the transmission, causes larger energy loss of the successor during its dwell time $\bar{\tau}_2$ so that the scattering probabilities of the two electrons are correlated, violating $\sqrt{P_0} +\sqrt{P_2} \le 1$  [Eqs.~\eqref{eq:P-HBT}-\eqref{eq:dE-HBT}, Fig.~\ref{fig:HBT_asQPC}]. \HS{This may explain the violation observed in Fig.~4 of Ref.~\cite{ubbelohde:partitioning}.
The marks (i), (ii), (iii) in Fig.~\ref{fig:HBT_asQPC}(b) 
correspond to the nonmonotonic features  (i), (ii), (iii) of Fig.~\ref{fig:HBT_asQPC}(a), respectively, in agreement with Ref.~\cite{ubbelohde:partitioning}. }

  
 
\begin{figure}[t]
 \includegraphics[width=\columnwidth]{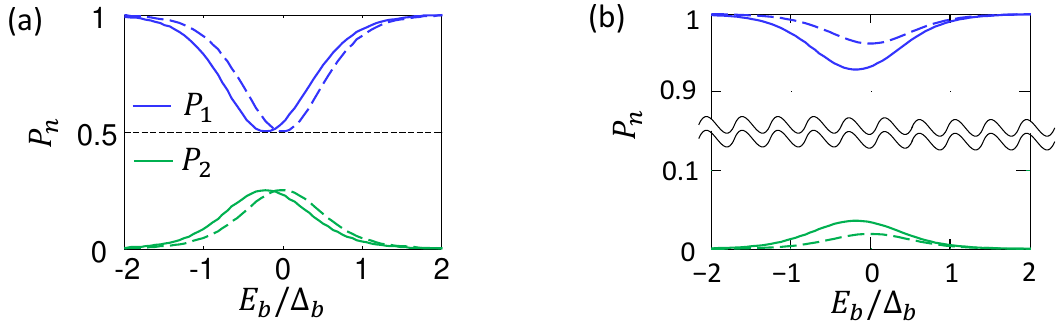}
 \caption{Partition probabilities $P_{n}$ by collision of two counterpropagating electrons at the symmetric constriction, as a function of $E_b$ in the noninteracting
(dashed curves) and interacting (solid) cases. $E_b$ is measured with respect to $\tilde{E}$. 
The electrons have (a) opposite spins or (b) same spins.
The same parameters with Fig.~\ref{fig:HBT_sQPC} are chosen.  
}
\label{fig:HOM}
\end{figure}

{\it Collision.---} We next consider two counterpropagating hot electrons that simultaneously arrive at the symmetric constriction [Fig.~\ref{fig:setup}(b)]. 
Their wave packets have the same Gaussian form
of mean energy $\tilde{E}$ at the barrier entrance.
Their spins are in a product state $\ket{\chi_1}\otimes \ket{\chi_2}$,
as generated by independent pumps.
In this case, the partition probabilities satisfy $P_0 = P_2$ and $P_1 = 1 - 2 P_2$. In Fig.~\ref{fig:HOM}, we compute $P_n$, using Eq.~\eqref{eq:Pdir}. 
The results qualitatively agree with the relations
\begin{equation}
  \label{eq:P-HOM-gen}
  \begin{aligned}
    P_2 \simeq
    &  |\mathbf{t}_{\tilde{E} + \delta E^{(\text{dir})} }|^2  | \mathbf{r}'_{\tilde{E} +\delta E^{(\text{dir})} }|^2   \\
  &- |\braket{\chi_1|\chi_2}|^2 |\mathbf{t}_{\tilde{E} +\delta E^{(\text{ex})} }|^2  | \mathbf{r}'_{\tilde{E} +\delta E^{(\text{ex})} }|^2
  \end{aligned}
\end{equation}
valid at $\sigma_E \ll \Delta_b$.
$\delta E^\text{(dir/ex)}$ is the kinetic energy change by direct/exchange interactions during the collision.



 
In Fig.~\ref{fig:HOM}(a) we consider electrons having opposite spins, $\braket{\chi_1|\chi_2}=0$.
In the noninteracting case, $\delta E^\text{(dir)}=0$
and the dependence of $P_2$ on $E_b$ has a peak of height $P_2 = 1/4$ at $E_b= \tilde{E}$ at which $|\mathbf{t}_{\tilde{E}}|^2= |\mathbf{r}'_{\tilde{E}}|^2 = 1/2$.
In the interacting case, $P_n$ is determined by $\tilde{E} + \delta E^\text{(dir)}$. $\delta E^\text{(dir)}$ is negative, as the distance between the electrons decreases in the collision. The peak of $P_2$ is shifted to lower $E_b$ by $|\delta E^{(\text{dir})}|$, but the peak height is still 1/4.

In Fig.~\ref{fig:HOM}(b) we consider electrons having the same spin, $\braket{\chi_1|\chi_2}=1$.
In the noninteracting case, the antibunching of $P_1 = 1$ and $P_2 = P_0 = 0$ happens in the plane wave limit of $\sigma_E = 0$. However deviation $P_2 \ne 0$ from the antibunching occurs at finite $\sigma_E / \Delta_b$,
where the form of the wave packet changes during barrier scattering~\cite{bellentani:coulomb}.
%
In the interacting case, further deviation happens, since 
$|\delta E^{(\text{ex})}|$ is smaller than $|\delta E^{(\text{dir})}|$ as usual.

\HS{Nonmonotonic behaviors of $P_n$, similar to those of Fig.~\ref{fig:HBT_asQPC}(a), can happen in the collision, when the electrons arrive at the barrier at different times more than $a_\textrm{cut}/v$.}


{\it Discussion.---} 
We develop a theory for scattering of two 
interacting electrons by a potential barrier, and emphasize kinetic energy change by their interaction during scattering delay times. The change depends on whether they copropagate or counterpropagate to approach the barrier, their relative arrival time at the barrier, and the non-monotonicity and asymmetry in the delay times.
For copropagating electrons, the preceding electron gains energy while the succeessor losses energy. For counterprogating cases, they both loss energy. The energy change results in nonmonotic dependence of their partition on the barrier height, correlation of their scattering probabilities, and reduction of fermionic antibunching in the collision.
Our finding provides a basic example of multiparticle scattering problems,
and will be useful in application of electron quantum optics to flying qubits, as combination of barrier partitioning and Coulomb interactions is essential for coupling multiple qubits.
  

Our finding does not rely on a specific form of the barrier and Coulomb potentials.
\HS{Our perturbative treatment of the Coulomb interaction is applicable when $\Gamma_\ell \bar{\tau} \sim \hbar \Gamma_\ell / \Delta_b \lesssim \Delta_b$; this condition is satisfied with usual constrictions~\cite{ubbelohde:partitioning} where $\bar{\tau} \sim$ subpicoseconds.}

We note that in the quantum Hall regime~\cite{feve:demand,dubois:minimal,jullien:quantum,bisognin2019quantum} where  
electron wave packets move along a quantum Hall edge channel, having low energy ($\leq$ 0.1 meV) close to the Fermi level, $\Gamma_\ell / \Delta_b$ may be so small that our effects are negligible.
When the scattering amplitudes of those packets at a quantum point contact are manipulated to be energy dependent (e.g., in nonequilibrium),
the scattering delay times will play a role, as in our study. It will be interesting to study interplay between the delay times and electron interactions of chiral Luttinger liquids along the edge.



\HS{We considered initial two-electron (anti-symmetrized) product states. This is supported by experiments~\cite{fletcher2019continuous} where the purity of electron states generated by a quantum-dot pump is very low. Nonetheless, entanglement in initial states can affect $P_n$, depending on its detailed form. For example, when two electrons are initially in an equal superposition of the initial states of Figs.~\ref{fig:HBT_sQPC}(b) and \ref{fig:HBT_sQPC}(e), their partition probability $P_n$ equals the average of the results in Figs.~\ref{fig:HBT_sQPC}(b) and \ref{fig:HBT_sQPC}(e). In this case, $P_n$ does not show the nonmonotonicity. Studies on the effects of general entanglement will be valuable.}
	
When the electrons occupy incoherent wave packets,
exchange interactions vanish, so their collision is governed by direct processes.
In this case, our theory is applicable~\cite{Fletcher:Park} also to the regime of $\Delta_b < \sigma_E$ with classical ensemble average, although it is developed for $\Delta_b > \sigma_E$.

It is known that in mesoscopic devices, scattering delay times play a role in nonlinear current response~\cite{pedersen:charge,christen1996gauge,sanchez2013scattering,dashti2021readout} at scatterers due to charge screening,  
although they are short as subpicoseconds ($\sim \hbar / \Delta_b$). 
Our work identifies their new role in multiparticle scattering.
This role was unnoticed in theories on two-particle scattering~\cite{Levinson:twoelectrons} including those for capacitively coupled conductors~\cite{goorden2007two}, numerical studies~\cite{bellentani:coulomb} for colliding electrons,
and classical descriptions~\cite{Pavlovska:classical}.
Note that the delay times differ from the traversal time~\cite{ryu:ultrafast,buttiker:traversal,landauer:barrier}.
 

We thank Jonathan Fletcher, Vyacheslavs Kashcheyevs, Masaya Kataoka, Wanki Park, and Niels Ubbelohde for discussions. This work is supported by Korea NRF via the SRC Center for Quantum Coherence in Condensed Matter (Grant No. 2016R1A5A1008184). SR acknowledges partial support from the Mar\'ia de Maeztu Program for Units of Excellence No. MDM2017-0711 funded by MCIN/AEI/10.13039/501100011033.



\clearpage
\newpage
\setcounter{page}{1}
\setcounter{equation}{0}
\setcounter{figure}{0}
\setcounter{secnumdepth}{2}
\renewcommand{\thefigure}{S\arabic{figure}}
\renewcommand{\theequation}{S\arabic{equation}}
\renewcommand{\bibnumfmt}[1]{[S#1]}
\renewcommand{\citenumfont}[1]{S#1}
\renewcommand{\thesection}{S\arabic{section}}  
\onecolumngrid
\begin{center}{\textbf{\large Supplemental Material: Partition of Two Interacting Electrons by a Potential Barrier}\\
    \vspace{2mm} 
    Sungguen Ryu$^{1,2}$ and H.-S. Sim$^{1,*}$}\\
    \vspace{2mm} 
    {\it $^1$Department of Physics, Korea Advanced Institute of Science and Technology, Daejeon 34141, Korea}\\
  \hbox{\it $^{2}$Institute for Cross-Disciplinary Physics and Complex Systems IFISC (UIB-CSIC), E-07122 Palma de Mallorca, Spain}
\end{center}
\vspace{1cm}

This material contains the parameters of the model, derivation of Eqs. (2)--(5), and saddle point constrictions.

\section{Interaction strength $W_0$}

We choose $W_0 = 144$ meV, based on $W_0 = q_e^2/(4\pi\epsilon_0 a_{\text{cut}})$,  the electron charge $q_e$, the vacuum permittivity $\epsilon_0$, and  $a_{\text{cut}}= 10$ nm ($\sim$ the magnetic length at 10 T).
The actual value of $W_0$ in experimental situation is yet unknown, and expected to be smaller than $W_0 = 144$ meV; it is determined by screening effects by the spatially separated two-dimensional Fermi sea and nearby gate electrodes. 
Our choice of $W_0 = 144$ meV is for demonstration of the nontrivial features in scattering of two interacting electrons by a potential barrier.
As discussed in the main text, our theory is applicable when $\Gamma_\ell \bar{\tau} \sim \hbar \Gamma_\ell / \Delta_b \lesssim \Delta_b $, since it is derived up to the lowest order of $W$; with our parameters including $W_0 = 144$ meV, $\Gamma_\ell \bar{\tau} \sim \Delta_b $.

\section{Derivation of Eq.~(2) of the main text}

We derive the two-particle partitioning probabilities $P_n$ in Eq.~(2) in the interaction picture. They are written as $P_n=   \lim_{t_f \rightarrow \infty}
  \braket{ \hat{P}_n(t_f)  }_{U_I(t_f) \ket{\Phi_0}}$.
Here $\ket{\Phi_0} \equiv (\ket{\phi_1\phi_2} \mp \ket{\phi_2\phi_1})/\sqrt{2} $ is the initial two-electron state with the sign $-$ (resp. $+$) for their spins in the spin triplet (resp. singlet).
We use the notation  $\ket{\psi\phi} = \ket{\psi}\otimes \ket{\phi}$.
$U_I(t_f)$ is the interaction-picture time evolution operator.
$\hat{P}_n(t_f)$ is the measurement operator projecting the state onto the subspace where $n$ electrons are in the lower ouput path at a long time $t_f$ after the barrier scattering,
\begin{equation}
  \label{eq:hatPn}
  \begin{aligned}
    \hat{P}_2(t_f) &= e^{i H_0 t_f} \int_\text{output path $d$} d x_d d x_d' \, \ket{x_d  x_d'} \bra{x_d x_d'} e^{-i H_0 t_f}\\
    \hat{P}_1(t_f) &= e^{i H_0 t_f}\Big[ \int_\text{output path $d$} d x_d  \int_\text{output path $u$} d x_u' \, \ket{x_d x_u'} \bra{x_d x_u'}
    +  \ket{x_u' x_d} \bra{x_u' x_d} \Big] e^{-i H_0 t_f}\\
    \hat{P}_0 (t_f)&= e^{i H_0 t_f} \int_\text{output path $u$} d x_u d x_u' \, \ket{x_u  x_u'} \bra{x_u x_u'} e^{-i H_0 t_f} ,
  \end{aligned}
\end{equation}
where $H_0$ is the Hamiltonian in the noninteracting situation. $\hat{P}_n(t_f)$ is symmetric under the particle exchange.

In the weak interaction, the time evolution operator is $U_I(t_f,0) \simeq 1 -i \int_0^{t_f} dt \, W_I(t)$ up to the first order of the interaction $W_I (t)= e^{i H_0 t} W e^{-i H_0 t}$.
We decompose the partitioning probability $P_n \simeq P_n^{(0)}  +\delta P_n^{\text{(dir)}} +\delta P_n^{\text{(ex)}} $ into the non-interacting value $P_n^{(0)}$ and the first-order corrections from the direct and exchange interaction
\begin{align}
  \delta P_n^{\text{(dir)}}
  &= \lim_{t_f \rightarrow \infty}  \int_0^{t_f} dt \braket{\phi_1 \phi_2|
    (-i) [ \hat{P}_n(t_f), W_I(t)] |\phi_1 \phi_2} , \label{eq:dP-dir} \\
  \delta P_n^{\text{(ex)}}
  &= \mp \lim_{t_f \rightarrow \infty} 
   \int_0^{t_f} dt\, \braket{\phi_1  \phi_2|
    ( -i)  [ \hat{P}_n(t_f),    W_I(t)   |\phi_2 \phi_1}. \label{eq:dP-exc} 
\end{align}
Note that $\lim_{t_f \rightarrow \infty} e^{i H_0 t}\hat{P_n}(t_f) e^{-i H_0t} =   \lim_{t_f \rightarrow \infty} \hat{P_n}(t_f)$ is satisfied.
This gives Eq.~(2).

\section{Wave-packet scattering state in Eq.~3}
\label{sec:psi}

We write the single-electron wave packet scattering state in Eq.~3 as a superposition of plane wave scattering states.
The scattering state $\psi_{\mathcal{E} u}$ (resp. $\psi_{\mathcal{E} d}$) is generated by the incoming plane wave of energy $\mathcal{E}$ in the upper (resp. lower) path. It is decomposed into the input path, output paths, and barrier region,
\begin{equation}
  \label{eq:psi-parts}
  \ket{\psi_{\mathcal{E} b}}
  = \ket{\psi^{\text{(in)}}_{\mathcal{E} b}}  +\sum_{a=u,d} \ket{\psi^{(\text{out},a)}_{\mathcal{E} b}}
  + \ket{\psi^{\text{(bar)}}_{\mathcal{E} b}}
  \qquad \qquad \text{ for } b=u, d.
\end{equation}
The wave function of $\psi^{(\text{in})}_{\mathcal{E} b}$ is  $\psi^{(\text{in})}_{\mathcal{E} b}(x_b) \sim  e^{i \mathcal{E} (x_{b}+L)/v }$ at position $x_{b} <-L$ and $\psi^{(\text{in})}_{\mathcal{E} b}(x_b) =0$ in the other region.  $\psi^{(\text{out},a)}_{\mathcal{E} b}$ is $s_{a b}(\mathcal{E}) e^{i \mathcal{E} (x_{a}-L)/v }$ at $x_{a} >L$ and 0 in the other region.
$\psi^{\text{(bar)}}_{\mathcal{E} b}$ is nonvanishing only inside the barrier $x_a, x_b \in [-L, L]$. Here we introduce the length $L$ of the barrier region for convenience. $L$ is large enough, so the region $[-L,L]$ includes the potential barrier; our results are insensitive to the value of $L$.
$s_{a b}(\mathcal{E})$ is the scattering matrix amplitude of the plane wave from the input path $b$ to the output path $a$.
We set $\hbar\equiv 1$ hereafter.

The initial wave packets $\phi_1$, $\phi_2$ are written as a superposition $ \ket{\phi_{m=1,2}} =  \int_0^\infty d \mathcal{E} \, \tilde{\phi}_m(\mathcal{E}) \ket{\psi_{\mathcal{E} \beta_m}}$ of plane-wave scattering states $\psi_{\mathcal{E} \beta_m}$ of energy $\mathcal{E}$.
$\beta_m$ denotes the input path of $\phi_m$.
$\beta_1 = \beta_2=d$ in the initially co-propagating case. $\beta_1=u$, $\beta_2=d$ in the HOM case.
Its time evolution $\ket{\phi_m(t)}$ is
\begin{equation}
  \label{eq:phi-t-exp}
  \ket{\phi_m(t)} = \int_0^\infty d \mathcal{E} \,
  \tilde{\phi}_m(\mathcal{E}) e^{-i \mathcal{E} t} \ket{\psi_{\mathcal{E} \beta_m}}
  \quad \text{for } m= 1,2.
\end{equation}
in the non-interacting case.
This is the state in Eq.~3. Using Eq.~\eqref{eq:psi-parts}, its wavefunction 
is found as
\begin{align}
    \braket{x| \phi_m^{(\text{in})}(t)}
    &= \int_0^\infty d \mathcal{E}\,
    \tilde{\phi}_m (\mathcal{E}) e^{-i \mathcal{E} (t-(x+L)/v)}
    \quad \quad  \quad  \quad \quad   \text{for } x \in \text{the input path } \beta_m  \label{eq:phi-t-exp-in-wavefun} \\
    \braket{x |\phi_m^{(\text{out},\alpha)}(t)}
    &= \int_0^\infty d \mathcal{E}\,
    \tilde{\phi}_m (\mathcal{E})   s_{\alpha \beta_m} (\mathcal{E}) e^{-i \mathcal{E} (t- (x-L)/v)}
    \quad \text{ for } x \in \text{the output path } \alpha. \label{eq:phi-t-exp-out-wavefun}
\end{align}


In an energy window $\mathcal{E} \in [E-\Delta E, E+\Delta E]$ with $\Delta E \ll \Delta_b$, $s_{a b}(\mathcal{E})$ depends on energy $\mathcal{E}$ approximately linearly,   
\begin{equation}
  \label{eq:approx-s}
 s_{a b}(\mathcal{E}) 
\simeq  s_{a b} (E) \Big[ 1+ \big\{ \frac{1}{2}\chi_{a b}(E)  + i\tau^{\text{D}}_{a b }(E)\big\} (\mathcal{E} -E)\Big].
\end{equation}
$\tau^{\text{D}}_{a b}(\mathcal{E})$ is the energy sensitivity $ \tau^{\text{D}}_{a b} (\mathcal{E}) \equiv \partial (\text{Im} \ln s_{a b})/\partial \mathcal{E} $ of the scattering phase shift.
$\chi_{a b}(\mathcal{E})$ is the energy sensitivity $\chi_{a b}(\mathcal{E}) \equiv |s_{a b}(\mathcal{E})|^{-2} (\partial  |s_{a b}|^2/\partial \mathcal{E})$ of the scattering probability at energy $\mathcal{E}$ for $a,b \in [u,d]$.
It satisfies 
\begin{equation}
  \label{eq:rel-chi}
  \sum_{a= u,d} |s_{a b}(\mathcal{E})|^2 \chi_{a b}(\mathcal{E}) = 0 ,
\end{equation}
which is obtained by differentiating the unitarity $\sum_{a=u,d} |s_{a b}(\mathcal{E})|^2 = 1$ with respect to $\mathcal{E}$.

\section{Derivation of Eq.~4}

We derive the probability of the single-electron wave packet to be in the barrier region in Eq.~4 of the main text,
\begin{equation}
  \label{eq:P-phi-barr}
  \braket{\phi_m^{(\text{bar})}(t)|\phi_m^{(\text{bar})}(t)} =
  v \sum_{\alpha=u,d} |s_{\alpha \beta_m} (E_m) |^2 \tau^{\text{D}}_{\alpha \beta_m} (E_m)| \braket{- L| \phi_m(t)}|^2.
\end{equation}
Here $A_m (t) = v | \langle - L | \phi_m (t) \rangle |^2$ is the arrival time distribution of $\phi_m(t)$ at the entrance of the barrier region.
A similar relation is found in Refs.~\cite{christen1996gauge-2,sanchez2013scattering-2,dashti2021readout-2} for 
plane-wave scattering states. For completeness, we derive the relation and then Eq.~4.

We consider two plane-wave scattering states $\ket{\psi_{\mathcal{E}b}}$ and $\ket{\psi_{\mathcal{E}'b}}$ coming from the same input path $b$ [see Eq.~\eqref{eq:psi-parts}]
and obtain their overlap  in the barrier,
\begin{equation}
  \label{eq:P-psi-barr}
  \braket{\psi^{\text{(bar)}}_{\mathcal{E} b}|\psi^{\text{(bar)}}_{\mathcal{E}' b}}
  =  -i v \sum_{a=u,d}  s^*_{ab} (\mathcal{E}) \frac{s_{ab} (\mathcal{E}) -s_{ab} (\mathcal{E}')}{\mathcal{E} -\mathcal{E}'}
 \quad \quad \quad \text{for } b= u,d.
\end{equation}
The overlap corresponds to the local density of states~\cite{christen1996gauge-2,sanchez2013scattering-2,dashti2021readout-2} in the barrier, 
which is related to charge oscillations in the barrier at frequency $(\mathcal{E} -\mathcal{E}')/h$~\cite{pedersen:charge-2}.
It is derived, using the conservation of probability flux of a superposition 
$  [ \ket{\psi_{\mathcal{E}b}}e^{-i \mathcal{E} t}
  + \ket{\psi_{\mathcal{E}'b}}e^{-i \mathcal{E}' t} ] / \sqrt{2v}$,
where the factor $1/\sqrt{v}$ makes the flux be 1.
We calculate the flux $J(t)$ of the superposition into the barrier (the net flux from the input path $b$ and output paths $a= u,d$).
The direct contribution of the superposition to the flux vanishes as $\sum_{a=u,d} |s_{ab}|^2=1$.
The interference contribution determines 
\begin{equation}
  \label{eq:J}
  J(t) =   \text{Re}\Big[e^{i (\mathcal{E} -\mathcal{E}')t}
  \Big\{1
  - \sum_{a=u,d}  s_{ab}^*(\mathcal{E}) s_{ab}(\mathcal{E}')   \Big\} \Big].
\end{equation}
And, the probability $p_{\text{bar}} (t)$ of finding the superposition in the barrier satisfies 
\begin{align}
  \frac{d p_{\text{bar}}}{d t} &=
   \frac{1}{v} \text{Re} \Big[ \braket{\psi^{\text{(bar)}}_{\mathcal{E} b}|\psi^{\text{(bar)}}_{\mathcal{E}' b}}
                                  i (\mathcal{E} -\mathcal{E}') e^{i (\mathcal{E}-\mathcal{E}')t} \Big].
 \label{eq:dP-barr}                                  
\end{align}
Using the flux conservation $J= dp_{\text{bar}}/dt$ and the unitarity $\sum_{a=u,d} |s_{ab}|^2=1$, we obtain Eq.~\eqref{eq:P-psi-barr}.
We further compute Eq.~\eqref{eq:P-psi-barr} when $|\mathcal{E} -\mathcal{E}'| \ll \Delta_b$. Under this condition, we approximate
$[s_{ab}(\mathcal{E}) -s_{ab}(\mathcal{E}')]/(\mathcal{E}-\mathcal{E}')
= d s_{ab}/d \mathcal{E}$ and apply Eq.~\eqref{eq:approx-s} to get
$d s_{ab}/d \mathcal{E} = s_{ab}(\mathcal{E}) ( \chi_{ab}(\mathcal{E})/2 + i \tau^{\text{D}}_{ab}(\mathcal{E}))$.
Combining this with Eq.~(\ref{eq:rel-chi}), 
we obtain
\begin{equation}
  \label{eq:ovl-psi-barr}
  \braket{\psi^{\text{(bar)}}_{\mathcal{E} b}|\psi^{\text{(bar)}}_{\mathcal{E}' b}}
  = v \sum_{a =u,d} |s_{ab}(\mathcal{E})|^2  \tau^{\text{D}}_{ab}(\mathcal{E})
  + \mathcal{O}\big(\frac{\mathcal{E}-\mathcal{E}'}{\Delta_b}\big)^2.
\end{equation}

Next, we derive the overlap, in the barrier, of two scattering states coming from different input paths $b$, $b'\neq b$,
\begin{equation}
  \label{eq:psi-br-off}
  \braket{\psi^{(\text{bar})}_{\mathcal{E} b} |\psi^{(\text{bar})}_{\mathcal{E}' b'}}
  = \sum_{a=u,d} s_{a b}^*(\mathcal{E}) s_{a b'}(\mathcal{E}')
  \braket{\psi^{(\text{in})}_{\mathcal{E} b} |\psi^{(\text{in})}_{\mathcal{E}' b} }. 
\end{equation}
It is found from the orthogonality $\braket{\psi_{\mathcal{E} b}|\psi_{\mathcal{E}' b'}}=0$ for $b\neq b'$ that implies
$\braket{\psi^{(\text{bar})}_{\mathcal{E} b}|\psi^{(\text{bar})}_{\mathcal{E}' b'}}
= -\sum_{a=u,d} \braket{\psi^{(\text{out},a)}_{\mathcal{E}b } |\psi^{(\text{out},a)}_{\mathcal{E}' b' }}
= -\sum_{a=u,d} s_{ab}^*(\mathcal{E}) s_{ab'}(\mathcal{E}') \int_L^\infty d x \, e^{i (\mathcal{E}'-\mathcal{E})(x-L)/v}$,
and the unitarity $\sum_{a} s_{ab}^*(\mathcal{E}) s_{ab'}(\mathcal{E})=0$.
Note that in Eq.~(\ref{eq:psi-br-off}), the terms  start from the first order of $(\mathcal{E}-\mathcal{E}')/\Delta_b$ due to the unitarity.
The properties of $\psi^{\text{(bar)}}_{\mathcal{E} b}$ in Eqs.~(\ref{eq:ovl-psi-barr}) and (\ref{eq:psi-br-off}) determine the effect of the delay time in the barrier on the partitioning probability of two interacting electrons when the Coulomb interaction is weak and $\sigma_E \ll \Delta_b$.

Finally, we obtain Eq.~4 when $\sigma_E \ll \Delta_b$.
We expand the overlap $\braket{\phi^{(\text{bar})}_{m}(t) |\phi^{(\text{bar})}_{m}(t)}$
into the scattering states, 
\begin{equation}
  \label{eq:ovl-phi-barr}
  \braket{\phi^{(\text{bar})}_{m}(t) |\phi^{(\text{bar})}_{m}(t)}  
  = \int_0^\infty d \mathcal{E} \int_0^\infty d \mathcal{E}'
  \tilde{\phi}^*_m(\mathcal{E}) \tilde{\phi}_m(\mathcal{E}')  e^{i (\mathcal{E} -\mathcal{E}')t}
   \braket{\psi^{\text{(bar)}}_{\mathcal{E} \beta_m}|\psi^{\text{(bar)}}_{\mathcal{E}' \beta_m}} ,
\end{equation}
and apply Eq.~(\ref{eq:ovl-psi-barr}) with substituting
$\sum_{a} |s_{ab}(\mathcal{E})|^2 \tau^{\text{D}}_{ab} (\mathcal{E}) \rightarrow
\sum_{a} |s_{ab}(E_m)|^2\tau^{\text{D}}_{ab} (E_m) $ (which is valid when $\sigma_E \ll \Delta_b$).
The remaining integral becomes $\int_0^\infty d \mathcal{E} d \mathcal{E}'
\phi^*_m(\mathcal{E}) \phi_m(\mathcal{E}')
e^{i (\mathcal{E}-\mathcal{E}') t} = v |\braket{-L|\phi_m(t)}|^2$. 
Combining these, we get Eq.~4.

\section{Derivation of Eq.~5}

We derive Eq.~5 of the main text.
For the purpose, it is useful to  introduce
a single-particle projection operator $\hat{p}_{\alpha}$ at an output path~$\alpha$
\begin{equation}
  \label{eq:hat-p}
  \hat{p}_{\alpha} \equiv \lim_{t_f \rightarrow \infty} u^\dagger(t_f)
  \int_L^\infty d x_{\alpha} \ket{x_\alpha}\bra{x_\alpha} u(t_f)  = \sum_{\mathcal{E}} \ket{\psi^{(-)}_{\mathcal{E} \alpha}} \bra{\psi^{(-)}_{\mathcal{E} \alpha}}  \quad \quad \text{for }\alpha = u, d.
\end{equation}
$u(t_f)$ is the single-electron time evolution operator in the noninteracting case. 
The scattering state $\psi^{(-)}_{\mathcal{E} \alpha}$ of energy $\mathcal{E}$ has a plane wave  outgoing  from the barrier only into the path $\alpha$, satisfying $\braket{\psi^{(-)}_{\mathcal{E} \alpha}| \psi_{\mathcal{E}' \beta}}= \delta_{\mathcal{E} \mathcal{E}'} s_{\alpha \beta}(\mathcal{E})$.
$\hat{p}_{\alpha}$ gives the probability that the electron eventually moves to the path $\alpha$ after barrier scattering. Using Eq.~\eqref{eq:psi-parts}, we find 
\begin{align}
 \hat{p}_{\alpha }  \ket{\psi_{\mathcal{E} \beta} } 
  &= \sum_{b=u,d } s^*_{\alpha b} (\mathcal{E})  s_{\alpha \beta}(\mathcal{E}) \ket{\psi_{\mathcal{E} b}}
= \sum_{b=u,d } s^*_{\alpha b} (\mathcal{E})  s_{\alpha \beta}(\mathcal{E}) \big[\ket{\psi^{(\text{in})}_{\mathcal{E} b}} +\ket{\psi^{(\text{bar})}_{\mathcal{E} b}} \big]
    + \ket{\psi^{(\text{out},\alpha )}_{\mathcal{E} \beta}}. \label{eq:hatP-2}
\end{align}
The first equality is found with $\mathbb{1} = \sum_{\mathcal{E}', b} \ket{\psi_{\mathcal{E}' b}^{(-)}} \bra{\psi_{\mathcal{E}' b}^{(-)}}$
to $\hat{p}_{\alpha }  \ket{\psi_{\mathcal{E} \beta} }$ and $\braket{\psi^{(-)}_{\mathcal{E} \alpha}| \psi_{\mathcal{E}' \beta}}= \delta_{\mathcal{E} \mathcal{E}'} s_{\alpha \beta}(\mathcal{E})$.
In the second,  we use Eq.~(\ref{eq:psi-parts}) and 
$ \sum_{a,b=u,d} s_{\alpha b}^* (\mathcal{E}) s_{\alpha \beta} (\mathcal{E})
    \int_L^\infty  d x_{a} s_{a b}(\mathcal{E}) e^{i \mathcal{E} (x_{a}-L)/v} \ket{x_a} 
    =  \ket{\psi_{\mathcal{E} \beta}^{(\text{out},\alpha)}}$.

In our perturbation approach where the Coulomb interaction between the two electrons is weak, it is convenient to use the following single-particle properties and single-particle conditional states.
We derive the properties of the projection operator $\hat{p}_\alpha$ for the wave packet $\phi_m$,
\begin{align}
  \braket{\hat{p}_{\alpha} }_{\ket{ \phi_{m}}}
  &= \int_0^\infty d \mathcal{E} \, |s_{\alpha \beta_m}(\mathcal{E})|^2 |\tilde{\phi}_m(\mathcal{E})|^2  
= |s_{\alpha \beta_m}(E_m)|^2 + \mathcal{O}((\sigma_E/\Delta_b)^2 ), \label{eq:exp-p-2} \\
  \braket{f(\hat{x}) \hat{p}_{\alpha}}_{\ket{ \phi_m(t)}}
  &=  \braket{\hat{p}_{\alpha} }_{\ket{ \phi_{m}}} \braket{ f(\hat{x})}_{\ket{\phi_{m|\alpha}(t)}}
    [1+ \mathcal{O}(\sigma_E/\Delta_b)], \label{eq:ovl-fp-dir} 
\end{align}
expanding $|\phi_m\rangle$ into scattering states and using Eq.~(\ref{eq:hatP-2}).
$\braket{\cdots}_{|\alpha \rangle}$ is the expectation value of $\cdots$ for the state $|\alpha \rangle$.
$E_{m}$ is the mean energy of $\phi_{m}$.
$f(\hat{x})$ is an arbitrary function of position operator $\hat{x}$.
$\phi_{m|\alpha}(t)$ is a conditional state of an electron that occupies $|\phi_m \rangle$ at the initial time $t=0$ and eventually moves into the output path $\alpha$, 
\begin{equation}
  \ket{\phi_{m| \alpha}(t)}
  \equiv \ket{\phi_m^{(\text{in})}(t)} +\ket{\phi_m^{(\text{bar})}(t)}
  + \frac{\ket{\phi_m^{(\text{out},\alpha)}(t)}}{s_{\alpha \beta_m }(E_m) }.  \label{eq:phi-cond}
\end{equation}
Its norm is $1 + O((\sigma_E/\Delta_b)^2)$ when $|\phi_m \rangle$ is Gaussian-like.
This conditional state is useful when the interaction is perturbatively treated.  
  It propagates in the output path $\alpha$, having the form almost identical to that of the initial wave packet but with the time delay $\tau^D_{\alpha \beta_m}(E_m)$; the difference between the forms is $O((\sigma_E/\Delta_b)^2)$.

We compute the correction of the partitioning probability, $\delta P_n^{(\text{dir})}$, by the direct Coulomb interaction in Eq.~2, up to the first order of $\sigma_E/\Delta_b$ and $\delta E/\Delta_b$, where $\delta E$ is the kinetic energy change by the Coulomb force 
during the barrier delay time.

To show the derivation, we use a notation $P^{\text{(dir)}}_{\alpha_1 \alpha_2}$ for the case that two initial wave packets $\phi_1$ and $\phi_2$ in input paths $\beta_1$, $\beta_2$ move to output paths $\alpha_1$, $\alpha_2$, respectively. It is related with $\delta P_n^{\text{(dir)}}$ in Eq.~\eqref{eq:dP-dir} as
$ P_2^{\text{(dir)}} = P^{\text{(dir)}}_{dd}$, $ P_0^{\text{(dir)}} = P^{\text{(dir)}}_{uu}$, $ P_1^{\text{(dir)}} =  P^{\text{(dir)}}_{du}+ P^{\text{(dir)}}_{ud}$.
In the weak interaction regime, we find $\delta P^{\text{(dir)}}_{\alpha_1 \alpha_2}
 = \sum_{m=1,2} \delta P^{(\text{dir},m)}_{\alpha_1 \alpha_2}$,  
\begin{align}
  \delta P_{\alpha_1 \alpha_2}^{(\text{dir},1)}
  &= \braket{\hat{p}_{\alpha_2}}_{\ket{\phi_2}} \int_0^\infty dt \, \braket{\phi_1(t)|(-i) [ \hat{p}_{\alpha_1}, w_1]|\phi_{1 }(t)},   \label{eq:Pdir-m} \\
  w_1 &\equiv  \braket{\phi_{{2|
        \alpha_{2}}}(t)|W |\phi_{2|\alpha_{2}}(t)} \label{eq:wm-def}  \\
  \delta P_{\alpha_1 \alpha_2}^{(\text{dir},2)}
  &= \braket{\hat{p}_{\alpha_{1}}}_{\ket{\phi_{1}}} \int_0^\infty dt \, \braket{\phi_2(t)|(-i) [ \hat{p}_{\alpha_2}, w_2]|\phi_{2 }(t)}, \nonumber \\
  w_2 &\equiv  \braket{\phi_{{1|
        \alpha_{1}}}(t)|W |\phi_{1|\alpha_{1}}(t)}. \nonumber
\end{align}
$\delta P^{(\text{dir},1)}_{\alpha_1 \alpha_2}$ is the change of the partitioning probability due to the interaction effect on the particle of index $1$ in the case that the other particle of index $2$ follows the unperturbed noninteracting evolution and eventually move to the output path $\alpha_{2}$. The unperturbed evolution is described by the conditional state $\ket{\phi_{2 | \alpha}(t)}$ in Eq.~(\ref{eq:phi-cond}), and 
$w_1 $ is the single-particle operator describing the direct Coulomb potential on the particle $1$.
$\delta P^{(\text{dir},2)}_{\alpha_1 \alpha_2}$ has the same meaning with $\delta P^{(\text{dir},1)}_{\alpha_1 \alpha_2}$, but with interchange of the particle indices $1 \leftrightarrow 2$.
Equation~(\ref{eq:Pdir-m}) is obtained by decomposing $\hat{P}_{\alpha_1 \alpha_2}$ into two single-particle projection operators $\hat{p}_{\alpha_m}$ in Eq.~(\ref{eq:hat-p}) and using
the expansion $[\hat{P}_{\alpha_1 \alpha_2}, W] = \hat{p}_{\alpha_1}\otimes \mathbb{1} [ \mathbb{1}\otimes \hat{p}_{\alpha_2}, W] + [\hat{p}_{\alpha_1}\otimes \mathbb{1}, W] \mathbb{1}\otimes \hat{p}_{\alpha_2} $ and  Eq.~(\ref{eq:ovl-fp-dir}).

We evaluate $\delta P^{(\text{dir},1)}_{\alpha_1 \alpha_2}$ 
\begin{align}
 \delta P^{(\text{dir},1)}_{\alpha_1 \alpha_2}
  &= \braket{\hat{p}_{\alpha_{2}} }_{\ket{\phi_{2}}} \big( \frac{\partial  |s_{\alpha_1 \beta_1}(\mathcal{E})|^2}{\partial \mathcal{E}}|_{\mathcal{E} = E_1}\big)
    \delta E^{(\text{dir})}_{1|\alpha_1 \alpha_2},  
    \label{eq:P-dir-m-2} \\
    \delta E^{(\text{dir})}_{1|\alpha_1 \alpha_2}
    &=\int_0^\infty dt\, \Big[ \braket{-v \frac{\partial W}{\partial \hat{x}_1}}_{\ket{\phi^{(\text{in})}_{1}(t)} \otimes \ket{\phi_{2|\alpha_{2}}(t)}}
     +  \bar{\tau}_1  A_1(t) \braket{-v \frac{\partial W}{\partial \hat{x}_1} }_{\ket{x=0}\otimes \ket{\phi_{2|\alpha_{2}}(t)}}\Big],
    \label{eq:dE-m-2}
\end{align}
using the scattering states and Eqs.~(\ref{eq:approx-s}), (\ref{eq:ovl-psi-barr}), (\ref{eq:psi-br-off}).
 $\delta P^{(\text{dir},2)}_{\alpha_1 \alpha_2}$ has the same form but with $1 \leftrightarrow 2$.
In Eq.~\eqref{eq:P-dir-m-2},  $\braket{\hat{p}_{\alpha_{2}} }_{\ket{\phi_{2}}}$ describes the probability for particle 2 moving to path $\alpha_2$ in the noninteracting case
and the other part $\big( \partial  |s_{\alpha_1 \beta_1}(\mathcal{E})|^2 / \partial \mathcal{E}|_{\mathcal{E} = E_1}\big)  \delta E^{(\text{dir})}_{1|\alpha_1 \alpha_2}$ describes the change of the scattering probability of particle 1 due to its energy change $\delta E^{(\text{dir})}_{1|\alpha_1 \alpha_2}$ by the direct interaction with particle 2.
In Eq.~(\ref{eq:dE-m-2}),
the first term means the power gain $-\partial W/\partial \hat{x}_1$ of particle $1$ by the Coulomb force  when particle 1 is in the input path and particle 2 is in the conditional state going to path $\alpha_2$. 
The second term means the power gain of particle 1  when particle 1 is inside the barrier  and  particle 2 is in the conditional state going to path $\alpha_2$.
In the second term, $ \bar{\tau}_1  A_1(t) = \braket{\phi_{1}^{\text{(bar)}} (t)|\phi_{1}^{\text{(bar)}}(t)}$ is the probability that particle 1 is inside the barrier as shown in Eq.~4.
The expressions in Eq.~\eqref{eq:dE-m-2} are independent of choice of the length $2L$ of the barrier region, when the length is sufficiently long so that the barrier region includes the region of tunneling of particles in the saddle constriction. The second term of Eq.~\eqref{eq:dE-m-2} gives $\Gamma^{\text{(dir)}}_\alpha$ in Eq.~5 of the main text.

In a similar way, we derive $\Gamma^{\text{(ex)}}$ in Eq.~5. We utilize $\tilde{\phi}_1(\mathcal{E})=\tilde{\phi}_2(\mathcal{E})$ and the following properties [cf.  Eqs.~(\ref{eq:exp-p-2}) and (\ref{eq:ovl-fp-dir})],
\begin{align}
    \braket{\phi_2| \hat{p}_{\alpha}|\phi_1 }
  &= \int_0^\infty d \mathcal{E} \,
    s_{\alpha \beta_2}^*(\mathcal{E}) s_{\alpha \beta_1}(\mathcal{E})
    \tilde{\phi}_2^*(\mathcal{E}) \tilde{\phi}_1 (\mathcal{E}) 
    = s_{\alpha \beta_2}^*(E_2) s_{\alpha \beta_1}(E_1)                                                               + \mathcal{O}((\sigma_E/\Delta_b)^2), \label{eq:ovl-p-2} \\
  \braket{ \phi_{2}(t) |f(\hat{x}) \hat{p}_{\alpha}| \phi_{1}(t) }  
  &=     \braket{\phi_2| \hat{p}_{\alpha}|\phi_1 }
    \braket{ \phi_{2| \alpha}(t)| f(\hat{x}) \mathcal{I}|  \phi_{1|\alpha}(t)}
    [1 + \mathcal{O}(\sigma_E/\Delta_b)], \label{eq:ovl-fp-exc} \\
  \mathcal{I} &\equiv \int_{-\infty}^{-L} dr  \, \ket{r_d=r}\bra{r_u=r} + \ket{r_u=r}\bra{r_d=r}. \label{eq:Inv}
\end{align}
$\mathcal{I}$ is the single-particle operator for the inversion in the input path subspace.
The partitioning probability becomes
\begin{align}
  \label{eq:dP-exc22}
  \delta P_2^{(\text{ex})}
  &= \mp 2 \text{Re}\Big[
    \braket{\phi_2| \hat{p}_{\alpha}|\phi_1 }  \int_0^\infty dt \,
\braket{\phi_{1}(t)|(-i)[\hat{p}_d, w_{\text{ex}}]|\phi_{2}(t)} \Big] , \\
  w_{\text{ex}}
  &\equiv \braket{\phi_{2|d}(t)|W \mathcal{I} |\phi_{1|d}(t)}  \label{eq:wexc-def}.
\end{align}
The sign $\mp$ comes from Eq.~\eqref{eq:dP-exc}.
$w_{\text{ex}}$ is a single-particle operator describing exchange interactions given that the particle providing the Coulomb force follows its noninteracting time evolution and goes to the lower output path.
Then
\begin{align}
  \delta  P^{(\text{ex})}_2
  & = \mp   \Big[  |s_{d \beta_1}(E_1)|^2  \big( \frac{\partial  |s_{d \beta_2} (\mathcal{E}) |^2}{\partial \mathcal{E}}|_{\mathcal{E} = E_2} \big) + |s_{d \beta_2}(E_2)|^2   \big( \frac{\partial  |s_{d \beta_1} (\mathcal{E}) |^2}{\partial \mathcal{E}}|_{\mathcal{E} = E_1} \big) \Big]
  \delta E^{(\text{ex})}    ,  \label{eq:dP-exc-2} \\
  \delta E^{(\text{ex})}
  &=\text{Re } \int_0^\infty dt \Big[ \braket{\phi_{1}^{(\text{in})}(t) \otimes \phi_{2|d}(t) |
    \big(-v \frac{\partial W}{\partial \hat{x_1}} \mathcal{I} \otimes \mathcal{I} \big)| \phi_{2}^{(\text{in})}(t) \otimes \phi_{1|d}(t)} \nonumber \\
  &\qquad \qquad  + \bar{\tau}  A_m(t)
    \braket{x=0 \otimes \phi_{2|d}(t) |
    \big(-v \frac{\partial W}{\partial \hat{x_1}} \mathcal{I} \otimes \mathcal{I} \big)| x=0 \otimes \phi_{1|d}(t)}  \Big] ,
         \label{eq:dE-exc-fin}
\end{align}
using the scattering states, Eqs.~(\ref{eq:approx-s}), (\ref{eq:ovl-psi-barr}), (\ref{eq:psi-br-off}), and Eq.~(\ref{eq:hatP-2}). 
Equation~\eqref{eq:dP-exc-2}  
describes the change of the scattering probability by the exchange interaction.
The first  (resp. second)  term  of Eq.~(\ref{eq:dE-exc-fin}) 
describes the Coulomb power gain of a particle in the input path  (resp. barrier), given that both the particles move out to the lower output path. The second term of Eq.~(\ref{eq:dE-exc-fin}) gives $\Gamma^{\text{(ex)}}$ in Eq.~5.
We note that the above derivation of the exchange contributions are for the electrons being initially in the same wave packet and arriving at the barrier at the same time. When there is time delay between the arrivals of the electrons or when the electrons are initially in different packets, the exchange contributions in Eqs. 5 and 6 have additional terms.

\section{Saddle point constriction}
\label{sec:delayTime}

We calculate the delay times of transmission and reflection events at a saddle point constriction potential $V_\text{sad}$,
\begin{equation}
  \label{eq:U-saddle}
  V_\text{sad}(x,y) =
  \begin{cases}
    E_b -\frac{1}{2} m \omega_{x\text{L}}^2 x^2 +\frac{1}{2} m \omega_{y\text{L}}^2 y^2 & \text{for } x<0
    \\
    E_b   -\frac{1}{2} m \omega_{x\text{R}}^2 x^2 +\frac{1}{2}m \omega_{y\text{R}}^2 y^2 & \text{for } x > 0.
  \end{cases}
\end{equation}
We below obtain analytic results for  the case of $\omega_{x\text{L}} \omega_{y\text{L}} = \omega_{x\text{R}} \omega_{y\text{R}}$.
This case contains a symmetric saddle point constriction of $\omega_{x \text{L}}=\omega_{x \text{R}}$ and $\omega_{y \text{L}}=\omega_{y \text{R}}$ and an asymmetric constriction of $\omega_{x \text{L}}\neq \omega_{x \text{R}}$ and $\omega_{y \text{L}}\neq \omega_{y \text{R}}$. 
 
\begin{figure}[b]
  \centering
  \includegraphics[width=0.8\textwidth]{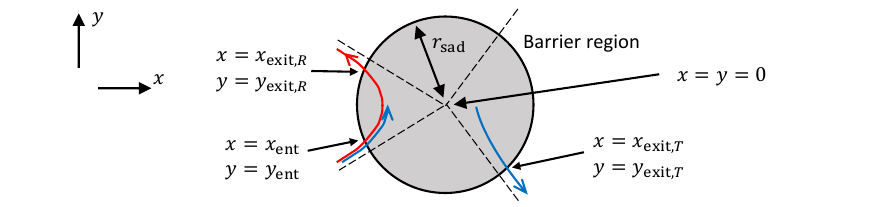}
  \caption{Saddle-point constriction. The saddle point is at $x=y=0$. The blue (red) arrow depicts the equipotential lines for the transmission (reflection) event. $(x_{\text{ent}}(\mathcal{E}) , y_{\text{ent}}(\mathcal{E}))$  is the entrance position of the line to the barrier, while 
$(x_{\text{exit,}T}(\mathcal{E}),  y_{\text{exit},T}(\mathcal{E}))$ and $(x_{\text{exit,}R}(\mathcal{E}),  y_{\text{exit,}R}(\mathcal{E}))$ are the exit positions for the transmission and reflection events, respectively.
}
  \label{fig:saddle}
\end{figure}

We use the definition~\cite{hauge:tunneling-2} of the delay time as the time during which the peak 
position of an 
incident wave packet of mean 
energy $\mathcal{E}$ (measured from the band bottom of the lowest Landau level; we consider an incident wave packet in the lowest Landau level~\cite{ryu:ultrafast-2}) enters and exits the barrier.
The delay time does not vanish only when the classical trajectory of an incident electron enters and exits the barrier.
Here we introduce  a circle of radius $r_{\text{sad}}$ from the saddle point, within which barrier scattering or tunneling mostly happens. See Fig.~\ref{fig:saddle}. We call the region inside the circle the barrier region. The radius is of the order of the magnetic length $l_B$~\cite{fertig:transmission-2}, where  $l_B \equiv \sqrt{\hbar/(m^*\omega_c)}$, $\omega_c$ is the cyclotron frequency, and $m^*$ is the effective electron mass. 
We choose $r_{\text{sad}}= 0.7 l_B$; the results in the main text does not depend on the specific choice. 
%

We apply the transformation in Ref.~\cite{fertig:transmission-2} to each regions of $x<0$ and $x>0$.
It maps the guiding-center motion of an electron in the two-dimensional saddle point potential in the strong magnetic field  into the problem of one-dimensional inverse harmonic barrier Hamiltonian $H_1=  E_b +\Delta_b (P^2 -X^2)$, where $\Delta_b = \omega_{x\text{L}}\omega_{y\text{L}}/(2 \omega_c)$. The cyclotron motion of the electron is described by another Hamiltonian which commutes with $H_1$; hence the transformation describes the situation where the electron remains in the lowest Landau level after barrier scattering.
The dimensionless coordinate $X$  of the guiding center relates with the original coordinate $x$,
\begin{equation}
  \label{eq:x-X}
  X =
  \begin{cases}
    \sqrt{\frac{\omega_{x\text{L}}}{\omega_{y\text{L}}}} \frac{x}{l_B}   & \text{ for } x<0
    \\
    \sqrt{\frac{\omega_{x\text{R}}}{\omega_{y\text{R}}}} \frac{x}{l_B} & \text{ for } x>0
    \end{cases}
\end{equation}
and $P$ is the canonical conjugate of $X$, namely $[X, P]=i$.
The transmission probability and delay time of the wave packet having energy $\mathcal{E}$ in the saddle point constriction are determined by $H_1$.
The transmission probability is $T(\mathcal{E})= 1/[1 + \exp[-\pi (\mathcal{E}-E_b)/\Delta_b]$.
The delay time $\tau^{\text{D}}_{T(R)}$  of the transmission (reflection) event can be written as
\begin{equation}
  \label{eq:tau-def}
    \tau^{\text{D}}_T(\mathcal{E}) = \lim_{X_0 \rightarrow \infty}
    [\tau_T^{\text{(long)}}(\mathcal{E} ) - \tau_T^{\text{(long-barr)}}(\mathcal{E} )], \quad \quad \quad \quad \quad 
    \tau^{\text{D}}_R (\mathcal{E}) =  \lim_{X_0 \rightarrow \infty}
    [\tau_R^{\text{(long)}}(\mathcal{E} ) -\tau_R^{\text{(long-barr)}}(\mathcal{E} )].
\end{equation}
$\tau^{\text{(long)}}_{T(R)}$ is the time for the packet to propagate from $X = -X_0$ to $X_0$ in the transmission (reflection) event. $X_0 \to \infty$ so that the domain $[-X_0, X_0]$ includes the barrier region sufficiently. $\tau^{\text{(long-barr)}}_{T}$ is the propagation time through domains $[-X_0, X_\textrm{ent} (\mathcal{E})]$ and $[X_{\textrm{exit},T} (\mathcal{E}),X_0]$, while $\tau^{\text{(long-barr)}}_{R}$ is the propagation time through domains $[-X_0, X_\textrm{ent} (\mathcal{E})]$ and $[X_{\textrm{exit},R} (\mathcal{E}), - X_0]$.
$X_{\text{ent}}(\mathcal{E})$ and $X_{\text{exit,}T(R)}(\mathcal{E})$ are related with the original coordinates $x_{\text{ent}}(\mathcal{E})$ and $x_{\text{exit,}T(R)}(\mathcal{E})$ at the intersection of the circle of radius $r_\textrm{sad}$ and the equipotential lines of $\mathcal{E}$ through Eq.~\eqref{eq:x-X},
\begin{equation}
  \label{eq:X-ent-exit}
  \begin{aligned}
    X_{\text{ent}} (\mathcal{E})
    &=  X_{\text{exit,}R} (\mathcal{E})
    =-\sqrt{\frac{\omega_{x\text{L}} \omega_{y\text{L}}}{\omega_{x\text{L}}^2 +\omega_{y\text{L}}^2}}
    \sqrt{ \big(\frac{r_{\text{sad}}}{l_B}\big)^2 - \frac{\omega_{x\text{L}}}{\omega_{y\text{L}}} \frac{\mathcal{E}-E_b}{\Delta_b}}
    \\ 
    X_{\text{exit,}T} (\mathcal{E}) &=  \sqrt{\frac{\omega_{x\text{R}} \omega_{y\text{R}}}{\omega_{x\text{R}}^2 +\omega_{y\text{R}}^2}}
    \sqrt{ \big(\frac{r_{\text{sad}}}{l_B}\big)^2 - \frac{\omega_{x\text{R}}}{\omega_{y\text{R}}}  \frac{\mathcal{E}-E_b}{\Delta_b}}.
  \end{aligned}
\end{equation}

 
$\tau^{\text{(long)}}_{T(R)}$ can be written in terms of the incoming (outgoing) part $\psi^{\text{(in)}}_{\mathcal{E}}$ ($\psi^{\text{(out)}}_{\mathcal{E}}$) of the scattering state, 
\begin{equation}
  \label{eq:tau-1st-def}
  \tau^{\text{(long)}}_T(\mathcal{E})  =
   \frac{\partial}{\partial \mathcal{E}} \big( \text{Im} \ln \frac{\psi^{\text{(out)}}_{\mathcal{E}}(X_0)}{\psi^{\text{(in)}}_{\mathcal{E}}(-X_0)} \big), \quad \quad \quad \quad \quad \quad\quad
  \tau^{\text{(long)}}_R(\mathcal{E})  =
    \frac{\partial}{\partial \mathcal{E}}\big( \text{Im} \ln \frac{\psi^{\text{(out)}}_{\mathcal{E}}(-X_0)}{\psi^{\text{(in)}}_{\mathcal{E}}(-X_0)}\big).
\end{equation}
Applying the assymptotic form~\cite{fertig:transmission-2} of the scattering state to $\psi^{\text{(out)}}_{\mathcal{E}}$ and  $\psi^{\text{(in)}}_{\mathcal{E}}$, we obtain
\begin{equation}
  \label{eq:tau-long}
  \tau^{\text{(long)}}_T(\mathcal{E})=  \tau^{\text{(long)}}_R(\mathcal{E})
  = \frac{1}{\Delta_b} \Big[ \frac{\ln 2}{2}  + \ln X_0
  - \frac{1}{2} \text{Re} \psi_\text{digamma} \big(\frac{1}{2}+ \frac{1}{2}i \big| \frac{\mathcal{E}-E_b}{\Delta_b} \big| \big) \Big].
\end{equation}
Here, $\psi_{\text{digamma}}$ is the digamma function, and $\Gamma(z) \Gamma(z+\frac{1}{2}) = \sqrt{2\pi} 2^{\frac{1}{2}-2 z} \Gamma(2 z)$ and  $ \Gamma(\frac{1}{4} + iy) \Gamma(\frac{3}{4} -i y) = \frac{\pi \sqrt{2} }{\text{cosh} (\pi y) + i \text{sinh}(\pi y) } $ were used. Note that in Eq.~(2.17) of Ref.~\cite{fertig:transmission}, $|X|^{(1/4)i \epsilon}$ should be replaced by $|X|^{-(1/4)i \epsilon}$;
this correction does not alter the result of the trasmission probability in Ref.~\cite{fertig:transmission-2}, but it is crucial in derivation of 
Eq.~\eqref{eq:tau-long}.
We compute the second terms of Eq.~\eqref{eq:tau-def}, using semiclassical approximations, 
\begin{equation}
  \label{eq:tau-term2-def}
    \tau_T^{\text{(long-barr)}}(\mathcal{E})
    =   \int_{-X_0}^{X_{\text{ent}}(\mathcal{E}) }      \frac{dX}{v(X; \mathcal{E})}
    + \int_{X_{\text{exit,}T}(\mathcal{E}) }^{X_0}   \frac{dX}{v(X; \mathcal{E})}, 
\quad \quad \quad \quad \quad
    \tau_R^{\text{(long-barr)}}(\mathcal{E})
    =  2\int_{-X_0}^{X_{\text{ent}}(\mathcal{E}) }     \frac{dX}{v(X; \mathcal{E})}.
\end{equation}
Here $v(X; \mathcal{E}) = 2 \Delta_b \sqrt{X^2+ (\mathcal{E}-E_b)/\Delta_b}$ is the semiclasical velocity at position $X$ and energy $\mathcal{E}$.
They become
\begin{equation}
  \label{eq:tau-long-barr}
  \begin{aligned}
    \tau_{T}^{\text{(long-barr)}}(\mathcal{E})
    &= \frac{1}{\Delta_b }\Big[\ln 2 +  \ln X_0
    - \frac{1}{2} \ln \frac{|\mathcal{E} -E_b|}{\Delta_b}
    + \frac{1}{2} g\big( \frac{X^2_{\text{ent}}(\mathcal{E}) }{ (\mathcal{E}-E_b)/\Delta_b} \big)
    + \frac{1}{2} g\big( \frac{X^2_{\text{exit},T}(\mathcal{E}) }{ (\mathcal{E}-E_b)/\Delta_b} \big)\Big] \\
   \tau_{R}^{\text{(long-barr)}}(\mathcal{E})
    &= \frac{1}{\Delta_b }\Big[\ln 2 +  \ln X_0
    - \frac{1}{2} \ln \frac{|\mathcal{E} -E_b|}{\Delta_b}
    + g\big( \frac{X^2_{\text{ent}}(\mathcal{E}) }{ (\mathcal{E}-E_b)/\Delta_b} \big) \Big].
  \end{aligned}
\end{equation}
Here $g(x) \equiv \ln |-\sqrt{|x|} + \sqrt{ |x| +\text{sgn}(x)}|$, where $\text{sgn}(x)$ is 1 when $x>0$ and -1 otherwise.
 Plugging Eqs.~\eqref{eq:tau-long} and ~\eqref{eq:tau-long-barr} into Eq.~\eqref{eq:tau-def}, we obtain the delay time 
\begin{equation}
  \label{eq:tau-result}
    \begin{aligned}
      \tau_T^{\text{D}}(\mathcal{E})
       &= \frac{1}{2\Delta_b} \Big[ \ln \big( \frac{|\mathcal{E} -E_b|}{2 \Delta_b}\big)
      -  \text{Re} \psi_\text{digamma} \big(\frac{1}{2}+ \frac{1}{2}i \big| \frac{\mathcal{E}-E_b}{\Delta_b} \big| \big)
      -  g\big( \frac{X^2_{\text{ent}}(\mathcal{E}) }{ (\mathcal{E}-E_b)/\Delta_b} \big)
      -  g\big( \frac{X^2_{\text{exit},T}(\mathcal{E}) }{ (\mathcal{E}-E_b)/\Delta_b} \big)\Big],
    \\
    \tau_R^{\text{D}}(\mathcal{E})
    &= \frac{1}{2 \Delta_b} \Big[ \ln \big( \frac{|\mathcal{E} -E_b|}{2 \Delta_b}\big)
      - \text{Re} \psi_\text{digamma} \big(\frac{1}{2}+ \frac{1}{2}i \big| \frac{\mathcal{E}-E_b}{\Delta_b} \big| \big)
      - 2 g\big( \frac{X^2_{\text{ent}}(\mathcal{E}) }{ (\mathcal{E}-E_b)/\Delta_b} \big) \Big].
  \end{aligned}
\end{equation}



We note that quantum tunneling or probabilistic scattering mostly happens in the region inside the circle of radius $r_\textrm{sad} \sim 0.7 l_B$, while semiclassical propagation occurs outside.
Hence the enhancement of the delay time for larger $r_\textrm{sad}$ is mostly due to the semiclassical propagation. We choose the smallest value of $r_\textrm{sad}$ with which there is not semiclassical propagation but quantum tunneling inside the barrier region.

\end{document}